\documentstyle[twocolumn,aps,psfig]{revtex}
\begin{document}             
\draft
\wideabs{          %Makes a two-column abstract    
\title{Determining the Wiedemann-Franz ratio from the thermal Hall conductivity: Application to %%@
Cu and $\rm YBa_2Cu_3O_{6.95}$.} 
\author{Y. Zhang$^1$, N.P. Ong$^1$, Z. A. Xu$^{1*}$, K. Krishana$^{1\dagger}$, R. Gagnon$^2$, and %%@
L. Taillefer$^3$}      
\address{
$^1$Joseph Henry Laboratories of Physics, Princeton University, Princeton, New Jersey 08544. \\
$^2$Department of Physics, McGill University, Montreal, Quebec, Canada H3A 2T8. \\
$^3$Department of Physics, University of Toronto, Toronto, Ontario, Canada M5S 1A7.
}
\date{\today}      % Deleting this command produces today's date.

%\newcommand{\ip}[2]{(#1, #2)}
                             % Defines \ip{arg1}{arg2} to mean
                             % (arg1, arg2).

\maketitle                   % Produces the title.

\begin{abstract}
The Wiedemann-Franz ($WF$) ratio compares the thermal and electrical conductivities in a metal.  %%@
We describe a new way to determine its value, based on the thermal Hall conductivity.  The %%@
technique is applied to copper and to untwinned YBaCuO. In the latter, we uncover a $T$-linear %%@
dependence and suppression of the Hall-channel $WF$ ratio. We discuss the implications of this %%@
suppression. The general suppression of the $WF$ ratio in systems with predominant electron-%%@
electron scattering is discussed.
\end{abstract}
}				%End of Wideabs command
The electron fluid in a metal is an excellent conductor of both electric charge and entropy. In %%@
familiar metals such as Au, Pb and Cu, the thermal conductivity of the electrons is so large that %%@
the phonons account for less than one percent of the heat current at any temperature $T$.  %%@
Wiedemann and Franz ($WF$) observed that the ratio of the thermal and electrical conductivities %%@
is very nearly the same in many metals \cite{Wiedemann}.  This ratio is expressed as the Lorenz %%@
number, which is conveniently written in dimensionless form as
\begin{equation}
{\cal L} = \frac{\kappa_e}{T\sigma} \left(\frac{e}{k_B}\right)^2,
\label{L}
\end{equation}
where $\kappa_e$ is the electronic thermal conductivity, $\sigma$ the electrical conductivity, %%@
$e$ the elementary charge, and $k_B$ is Boltzmann's constant.  In standard transport theory, %%@
${\cal L}$ equals the Sommerfeld value $\pi^2/3$ if the mean-free-paths ($mfp$) are assumed %%@
identical for charge and heat transport. Indeed, in all conventional metals, the observed $\cal %%@
L$ is rather close to this number above 273 K \cite{Ashcroft}.  However, below $\sim$273 K, the %%@
observed $\cal L$ falls significantly below this value, implying that the heat current is more %%@
strongly scattered relative to the charge current.  Because $\cal L$ compares directly the charge %%@
and entropy currents, it has contributed strongly to our current understanding of how charge and %%@
entropy currents are affected by distinct scattering processes in conventional metals %%@
\cite{Ziman}.

In solids, the observed thermal conductivity $\kappa$ is the sum of the electronic and phonon %%@
conductivity $\kappa_{ph}$, viz. $\kappa = \kappa_e + \kappa_{ph}$.  In conventional metals where %%@
$\kappa_{ph}\ll \kappa_e$, one may use $\kappa$ instead of $\kappa_e$ to evaluate $\cal L$ with %%@
negligible error. However, in conductors with relatively small carrier densities, $\kappa_{ph}$ %%@
is often much larger than $\kappa_e$.  These conductors (with resistivities $\rho$ exceeding 100 %%@
$\mu\Omega$cm) include many interesting conductors such as the cuprates, doped fullerenes, %%@
quasicrystals, and newer materials such as $\rm CaB_6$ \cite{CaB}.  Recent theoretical %%@
discussions of charge versus entropy currents are most applicable to these conductors.  %%@
Ironically, their $WF$ ratio seems experimentally inaccessible using conventional techniques. %%@
Hence, a fresh experimental approach is desirable.

The thermal Hall effect provides a rather efficient way to screen out the phonon heat current %%@
(even when it is dominant).  In zero field, the observed thermal current ${\bf J}_Q$ is parallel %%@
to the temperature gradient $-{\bf \nabla}T$ (applied $\parallel \hat{\bf x}$). A field $\bf %%@
H\parallel \hat{z}$ generates a Lorentz force that acts only on the electronic component of ${\bf %%@
J}_Q$.  This produces a transverse Hall current parallel to $\pm\bf \hat{y}$ (depending on the %%@
sign of the charge carriers).  Thus, the transverse Hall conductivity $\kappa_{xy}$ involves only %%@
the electrons (the phonon current is strictly unaffected by $\bf H$).  By forming the ratio in %%@
Eq. \ref{L} with $\kappa_{xy}$ and the electrical Hall conductivity $\sigma_{xy}$, we may measure %%@
directly the `Hall' Lorenz number ${\cal L}_{xy}\equiv [\kappa_{xy}/T\sigma_{xy}](e/k_B)^2$.  At %%@
sufficiently high $T$, ${\cal L}_{xy}$ approaches the Sommerfeld value \cite{Boltzmann}.

We have applied the technique to a detwinned crystal of optimally-doped cuprate $\rm %%@
YBa_2Cu_3O_{6.95}$ (YBCO) with a critical temperature $T_c$ = 93.3 K, and to elemental Cu.  A %%@
longitudinal gradient $-\partial_x T$ ($\approx 2$ K) is applied $\parallel\bf a$ of the YBCO %%@
crystal, and the transverse gradient $-\partial_y T$ is measured using a pair of thermocouples %%@
(chromel-constantan) versus field ($\bf H\parallel \pm c$).  The Hall signal $-\partial_H T$ is %%@
the field-{\em odd} component of $-\partial_y T$ \cite{expt}.  $\kappa_{xy}$ is computed as %%@
$[\partial_H T/\partial_x T](\kappa_{xx}\kappa_b/\kappa_a)$.  We measured $\kappa_a = %%@
\kappa_{xx}$, and used the anisotropy $\kappa_b/\kappa_a$ previously measured in another crystal %%@
from the same batch (see Ref. \cite{Gagnon}).  The copper sample was cut from oxygen-free high-%%@
conductivity (OFHC) stock, but was not vacuum annealed to further reduce lattice disorder. 

\begin{figure}[h]
\centerline{\psfig{figure=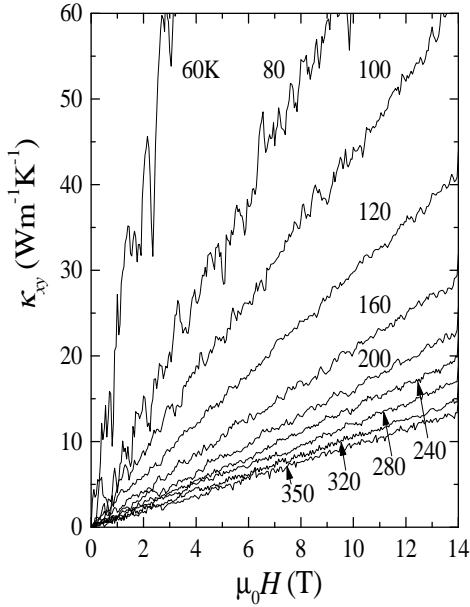,height=3.2in,width=2.7in}}
\caption{The field dependence of $\kappa_{xy}$ (electron-like) in unannealed OFHC copper at %%@
selected $T$ in Sample 1 (dimensions 4 mm $\times $ 0.3 mm $\times $ 50 $\mu$m). A gradient %%@
$\delta_x T\sim$0.4 K is applied along the sample length ($\parallel \bf \hat{x}$).  In the %%@
lowest curve (350 K), the value of $\kappa_{xy}$ at 14 T corresponds to a `Hall' temperature %%@
difference $\delta_y T$ across the sample width of $\sim 1.6$ mK.}
\label{KCu}
\end{figure}

First, we discuss the $T$-dependence of $\kappa_{xy}$ in Cu, and compare ${\cal L}_{xy}$ with the %%@
standard Lorenz number (see Ref. \cite{Borelius} for early data on $\kappa_{xy}$).  Figure %%@
\ref{KCu} displays traces of $\kappa_{xy}$ versus $H$ between 60 and 350 K measured in Sample 1.  %%@
In order to calculate $\cal L$ and ${\cal L}_{xy}$, we have also measured its electrical %%@
conductivity $\sigma$ and Hall conductivity $\sigma_{xy}$.  (Below 50 K, where $\kappa_{xy}$ and %%@
$\sigma_{xy}$ display increasing curvature vs. $H$, ${\cal L}_{xy}$ is obtained from their values %%@
in the limit of weak-fields.) 

The values of $\kappa$ vs. $T$ in Sample 1 are in close agreement with a compilation by Powell et %%@
al. \cite{Powell}.  However, in samples that have been carefully annealed, the peak value (at 20 %%@
K) may be twice as large [see Berman and MacDonald (BM)\cite{Berman}].

\begin{figure}[h]
\centerline{\psfig{figure=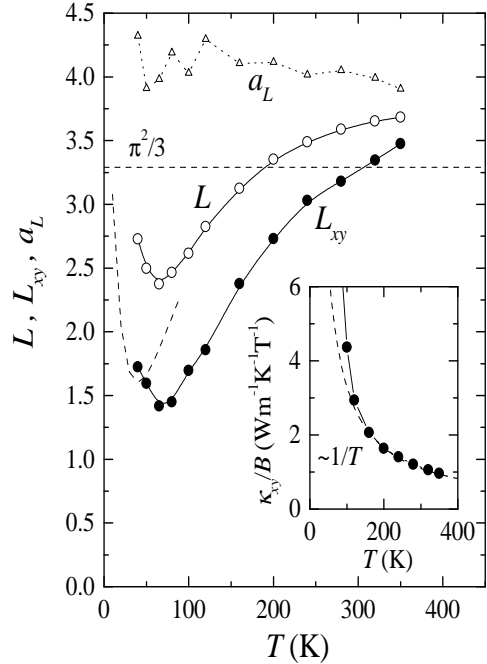,height=3.5in,width=2.8in}}
\caption{(Main Panel) The $T$ dependence of the conventional Lorenz number $\cal L$ (open) and %%@
Hall-Lorenz number ${\cal L}_{xy}$ (solid circles) measured in OFHC copper (Sample 1).  Both %%@
$\cal L$ and ${\cal L}_{xy}$ decrease from their high-$T$ asymptotic values ($\sim 10\%$ higher %%@
than the Sommerfeld value), with the latter falling faster.  To compare the two rates, we plot %%@
(open triangles) the quantity $a_L = {\cal L}^2/{\cal L}_{xy}$.  The near constancy of $a_L$ %%@
implies that ${\cal L}_{xy}\sim {\cal L}^2$ (see text).  The inset shows the $T$ dependence of %%@
$\kappa_{xy}/B$ measured in the limit $B\rightarrow 0$.  The dash line varies as $1/T$.}
\label{LCu}
\end{figure}

Figure \ref{LCu} shows that $\cal L$ and ${\cal L}_{xy}$ in our sample start at values slightly %%@
higher than the Sommerfeld value at 350 K, and decrease with falling $T$, with ${\cal L}_{xy}$ %%@
decreasing slightly faster. In Cu, the Debye temperature $\theta_D$ is 343 K.  Both Lorenz %%@
numbers reach a minimum near 60 K, and increase again at lower $T$. For comparison, we show %%@
(broken line) $\cal L$ reported by BM\cite{Berman}.  The minimum is deeper and occurs at a lower %%@
$T$ (20 K).  As discussed later, this is consistent with its higher purity.  

In conventional metals, electronic currents are limited by scattering of electrons by phonons (at %%@
finite $T$).  Large-angle scattering involving phonons with large wavevectors $\bf q$ ($q > k_F$, %%@
the Fermi wavevector) are equally disruptive of the charge and heat currents.  By contrast, %%@
small-angle scattering ($q\ll k_F$) relaxes only the heat current, leaving the charge current %%@
relatively unaffected \cite{Ziman}.  In terms of $\ell_S$ and $\ell_e$ (the mean-free-paths for %%@
entropy and charge transport, respectively) we may express $\cal L$ as $(\pi^2/3)\langle %%@
\ell_S\rangle/\langle\ell_e\rangle$, where $\langle\cdots\rangle$ denotes averaging over the %%@
Fermi Surface $FS$.  At high temperature ($T\geq \Theta_D$), we have $\ell_S\sim\ell_e$ because %%@
large-angle scattering is dominant.  As $T$ decreases below $\Theta_D$, the phonon population is %%@
increasingly skewed towards the small-$q$ limit, so that $\ell_e$ increases relatively faster %%@
than $\ell_S$.   

This accounts for the decrease in $\cal L$ in Fig. \ref{LCu}.  At very low $T$, when elastic %%@
scattering from impurities dominates, $\ell_e$ and $\ell_S$ are again equal, and $\cal L$ returns %%@
to the Sommerfeld value.  In our sample, this turnaround occurs near 60 K, whereas in the cleaner %%@
sample of BM, it occurs at 20 K.  
\noindent
In comparing $\cal L$ with ${\cal L}_{xy}$, we find that both display a minimum at nearly the %%@
same $T$.  However, on the high-$T$ side, ${\cal L}_{xy}$ falls faster with decreasing $T$.  The %%@
two rates are compared via $a_L\equiv {\cal L}^2/{\cal L}_{xy}$.  The near constancy of $a_L$ is %%@
consistent with the expectation \cite{Boltzmann} that ${\cal L}\sim \langle %%@
\ell_S\rangle/\langle\ell_e\rangle$, whereas ${\cal L}_{xy}\sim \langle %%@
\ell_S\rangle^2/\langle\ell_e\rangle^2$. Hence, ${\cal L}_{xy}$ falls faster because it involves %%@
the squares of the $mfp$'s.  In Cu, the Hall-Lorenz number approaches the same $WF$ ratio at high %%@
$T$, so that it provides information comparable to $\cal L$.  Moreover, additional information on %%@
scattering processes may be derived from its low $T$ behavior.

Thermal conductivity measurements have produced a wealth of information on the superconducting %%@
state of the cuprates.  In YBCO, they provided early evidence favoring an unconventional pairing %%@
symmetry \cite{Yu,Yu2}, and long lifetimes for the quasiparticles at low temperatures %%@
\cite{Krishana}.  The contributions of the chains to $\kappa$ \cite{Gagnon}, and the effect of %%@
hole doping \cite{Nakamura} also have been investigated.  Above $T_c$, however, the problem of %%@
estimating $\kappa_{ph}$ has been a serious obstacle to the extraction of $\kappa_e$.  In fact, %%@
neither the magnitude nor the $T$ dependence of $\kappa_e$ in the normal state may be regarded as %%@
experimentally established.  Hence, the present technique is especially appropriate.  While the %%@
behavior of $\kappa_{xy}$ in YBCO was investigated previously \cite{Krishana,Krishana2,Zeini} in %%@
the superconducting state, the rapid attenuation of the thermal Hall signal precluded %%@
measurements above $\sim$100 K.  The improved resolution now allows $\kappa_{xy}$ to be %%@
determined reliably up to 320 K.

\begin{figure}[h]
\centerline{\psfig{figure=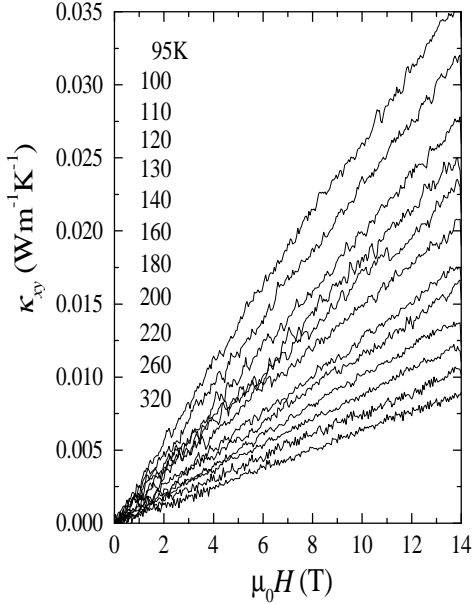,height=3.2in,width=2.7in}}
\caption{The field dependence of $\kappa_{xy}$ (hole-like) measured in untwinned $\rm %%@
YBa_{2}Cu_{3}O_{6.95}$ at selected $T$ (with $\bf H\parallel c$).  Each trace is in sequence with %%@
values of $T$ indicated.  At 320 K and 14 T, the value of $\kappa_{xy}$ corresponds to a Hall %%@
signal of $1.5$ mK (in an applied gradient $\delta_x T\sim 2 K$).}
\label{KYBCO}
\end{figure}
\begin{figure}[h]
\centerline{\psfig{figure=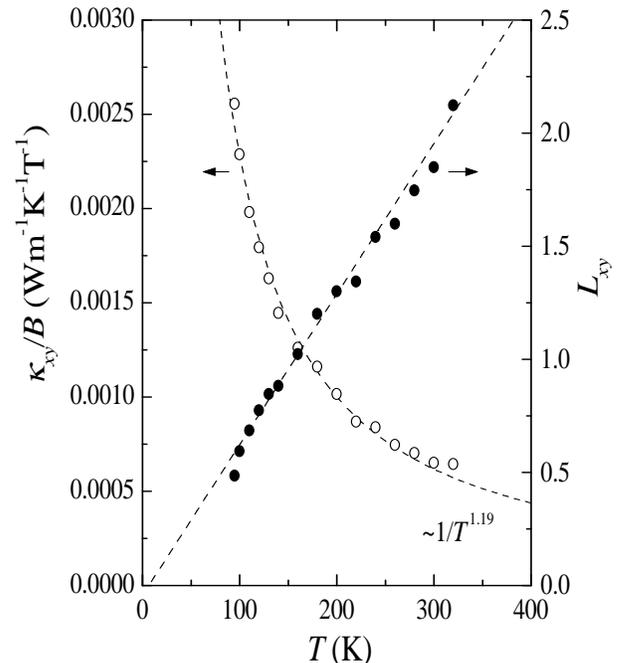,height=3.5in,width=3.2in}}
\caption{The $T$ dependence of the Hall-Lorenz number ${\cal L}_{xy}$ (solid circles) in %%@
untwinned $\rm YBa_{2}Cu_{3}O_{6.95}$ determined from $\kappa_{xy}/T\sigma_{xy}$ ($\cal L$ is %%@
unavailable for comparison).  Linear extrapolation of ${\cal L}_{xy}$ (broken line) shows that it %%@
attains the Sommerfeld value $\pi^2/3$ near 500 K.  The open circles represent the weak-field %%@
values of $\kappa_{xy}/B$ derived from Fig. \ref{KYBCO} (fit to curve $1/T^{1.19}$ is shown).}
\label{LYBCO}
\end{figure}

\noindent
Figure \ref{KYBCO} displays the $H$-dependence of $\kappa_{xy}$ at selected temperatures from 95 %%@
to 320 K.  It is worth a second remark that, as $\kappa_{xy}$ derives no contribution from the %%@
phonons, the raw experimental curves directly mirror the electronic heat current.  We proceed to %%@
compare it with the charge current.  In the normal state of YBCO, $\sigma_{xy}$ is known to vary %%@
as $1/T^3$.  This produces the anomalous $T$ dependence of the Hall coefficient $R_H$ that is so %%@
characteristic of the cuprates \cite{Chien}.  It is interesting to ask if the same dependence is %%@
observed in $\kappa_{xy}/T$ (dividing by $T$ to remove the heat capacity contribution).  As shown %%@
in Fig. \ref{LYBCO} (open symbols), the $T$ depedence of the $\kappa_{xy}/B$ is well-fitted to %%@
$T^{-1.2}$.  Hence, we find that $\kappa_{xy}/T$ actually has a weaker dependence than %%@
$\sigma_{xy}$.  

When we calculate ${\cal L}_{xy}$, we find that it varies linearly with $T$ (solid circles in %%@
Fig. \ref{LYBCO}).  In addition to this unusual $T$ dependence, its value from 95 to 320 K is %%@
significantly smaller than the Sommerfeld value.  

In some previous reports, values of $\cal L$ (often close to the Sommerfeld value) were obtained %%@
using a variety of arguments to subtract $\kappa_{ph}$ from $\kappa$.  However, these arguments %%@
are suspect for the following reasons.  In the cuprates, $\kappa_{ph}$ is strongly decreased by %%@
lattice disorder associated with dopants ($\kappa$ is 2 to 4 times larger in the pristine parent %%@
compound $\rm La_{2}CuO_4$ compared with doped compounds \cite{Nakamura}).  Moreover, within the %%@
plane, $\kappa_{ph}$ is strongly anisotropic.  In both $\rm YBa_2Cu_3O_7$ and its two-chain %%@
variant $\rm YBa_2Cu_4O_8$, $\kappa_{ph}$ is significantly larger along $\bf b$ (the chain axis) %%@
compared with $\bf a$, because of the anisotropy in elastic constants \cite{Zhang}.  These %%@
factors make comparisons between doped and undoped crystals, or between $\kappa$ measured along %%@
the axes $\bf a$ and $\bf b$ in the same sample, quite unreliable.  A different approach is to %%@
extrapolate the value of $\kappa_e$ of quasiparticles in the superconducting state to %%@
temperatures just above $T_c$.  A recent experiment \cite{Krishana2} reports extrapolated values %%@
of $\kappa_e$ in the range 0.8-1.0 W/mK at $T_c$.  With $\rho_a$ = 100 $\rm \mu\Omega cm$, this %%@
gives $\cal L$ = 1.1-1.4, which is closer to our ${\cal L}_{xy}$ than to the Sommerfeld value. 

Aside from the overall suppressed scale, the nominally $T$-linear dependence of ${\cal L}_{xy}$ %%@
is also unusual.  An extrapolation (broken line in Fig. \ref{LYBCO}) shows that it intersects the %%@
Sommerfeld value near 500 K.  We note that, the acoustic phonons in YBCO have a maximum energy of %%@
20 mv.  Above 200 K, the dominant acoustic phonons should have sufficient momenta to span the %%@
full $FS$ ($\Theta_D$ = 420 K is higher because of the large unit cell).  Certainly, by 320 K, we %%@
should have $\langle\ell_S\rangle = \langle\ell_e\rangle$ if electron-phonon scattering is the %%@
dominant mechanism for relaxing the two currents.  Hence, the small values of ${\cal L}_{xy}$ and %%@
the high temperature scale at which it attains the Sommerfeld value (500 K) seem incompatible %%@
with dominant electron-phonon scattering.  

In contrast, a suppressed $WF$ ratio may be expected in systems with dominant electron-electron %%@
($ee$) scattering.  A discussion of this point illustrates how normal ($N$) and Umklapp ($U$) %%@
scattering processes influence the $WF$ ratio.  As in the case of lattice thermal conduction %%@
\cite{Peierls,Callaway}, $N$ processes leave the total momentum of the electron gas unchanged, so %%@
that the charge current cannot relax without $U$ processes.  However, (unlike lattice conduction) %%@
$N$-process $ee$ scattering {\em does} relax the heat current because it causes a redistribution %%@
of energy between hot and cold electrons \cite{Peltier}.  This distinction implies that systems %%@
in which $ee$ scattering is dominant have a strongly reduced Lorenz number.  Moreover, because %%@
the relative weights of $N$ versus $U$ processes are not determined by $\Theta_D$ in $ee$ %%@
scattering, this reduction could prevail to very high $T$.  These issues are intimately related %%@
to behavior of the $WF$ ratio.  Hence, the $T$ dependences of $\kappa_{xy}$ and ${\cal L}_{xy}$ %%@
displayed in Fig. \ref{LYBCO} should place strong constraints on transport models for YBCO.

In summary, we have described measurements of the Hall-Lorenz number, obtained as the ratio of %%@
$\kappa_{xy}/T$ to $\sigma_{xy}$.  In Cu, this method gives results comparable to direct %%@
measurements of $\cal L$.  However, in systems in which lattice conduction is not small, the %%@
Hall-Lorenz experiment provides a direct comparison of the heat and charge currents of the charge %%@
carriers.  A quantitative understanding of the information derived from ${\cal L}_{xy}$ awaits %%@
comparison with microscopic calculations.  In view of the increased interest in systems with %%@
dominant $ee$ scattering, we expect high-temperature $WF$ ratio to play an increasingly prominent %%@
role.  

We acknowledge useful discussions with P. W. Anderson, B. Keimer, K. Damle, 
and S. Uchida.  N.P.O. and K.K. acknowledge support by the U.S. Office of Naval Research. 
N.P.O. acknowledges an award from the International Joint-Research grants from the New Energy and %%@
Industrial Tech. Develop. Org. (NEDO).  Y.Z. and Z.X. were supported by funds from a DMR-MRSEC %%@
award (NSF DMR 9809483) from the U.S. National Science Foundation.
\newline\noindent
$^*${\em Permanent address of Z.A.X.: Department of Physics, Zhejiang University, Hangzhou, %%@
China}\newline
$^\dagger${\em Present address of K.K.: Division of Engineering and Applied Sciences, Harvard %%@
Universiy, Cambridge MA 02138}

% now the references. delete or change fake bibitem. 

%

\begin{references}
\bibitem{expt} To attain the resolution used here ($\partial_y T\sim\pm$0.3 mK at 100 K), the %%@
long-term drift was kept below $\pm 3$ mK during the 3 hours of the field sweep. At some $T$, we %%@
verified that sweeps from $+14\rightarrow-14 T$ agreed with sweeps in the opposite sequence ($-%%@
14\rightarrow +14 T$).  The linearity of $\partial_y T$ vs. heat current was also verified.
\end{references}
\end{document}